\documentclass[aps,prl,reprint,superscriptaddress]{revtex4-1}

\usepackage{graphicx}

\begin{document}

\title{Drastic change of the Fermi surface across the metamagnetic transition in CeRh\(_2\)Si\(_2\)}

\author{K.~G\"{o}tze}
\affiliation{Hochfeld-Magnetlabor Dresden (HLD-EMFL), Helmholtz-Zentrum Dresden-Rossendorf and TU Dresden, D-01314 Dresden, Germany}

\author{D.~Aoki}
\affiliation{IMR, Tohoku University, Oarai, Ibaraki 311-1313, Japan}
\affiliation{INAC/PHELIQS, CEA Grenoble, F-38054 Grenoble, France}

\author{F.~L\'{e}vy-Bertrand}
\affiliation{Universit\'{e} Grenoble Alpes, Institut N\'{e}el, F-38000 Grenoble, France}
\affiliation{CNRS, Institut N\'{e}el, F-38000 Grenoble, France}

\author{H.~Harima}
\affiliation{Graduate School of Science, Kobe University, Kobe 657-8501, Japan}

\author{I.~Sheikin}
\email[Corresponding author.\\]{ilya.sheikin@lncmi.cnrs.fr}
\affiliation{Laboratoire National des Champs Magn\'{e}tiques Intenses (LNCMI-EMFL), CNRS, UGA, F-38042 Grenoble, France}
\affiliation{ICC-IMR, Tohoku University, Oarai, Ibaraki 311-1313, Japan}

\date{\today}

\begin{abstract}
We report high field de Haas-van Alphen (dHvA) effect measurements in CeRh$_2$Si$_2$ both below and above the first-order 26 T metamagnetic transition from an antiferromagnetic to a polarized paramagnetic state. The dHvA frequencies observed above the transition are much higher than those observed below, implying a drastic change of the Fermi-surface size. The dHvA frequencies above the transition and their angular dependence are in good agreement with band-structure calculations for LaRh$_2$Si$_2$, which correspond to CeRh$_2$Si$_2$ with localized $f$ electrons. Given that the $f$ electrons are also localized at low fields in CeRh$_2$Si$_2$, the Fermi-surface reconstruction is due to the suppression of antiferromagnetism and the restoration of the crystallographic Brillouin zone rather than the delocalization of the $f$ electrons. This example suggests that the intuitive notation of ``small'' and ``large'' Fermi surfaces commonly used for localized and itinerant $f$ electrons, respectively, requires careful consideration, taking into account the modification of the Brillouin zone in the antiferromagnetic state, when interpreting experimental results.
\end{abstract}

\pacs{}

\maketitle

Quantum critical points (QCPs), i.e., continuous phase transitions at zero temperature, play a key role in the physics of heavy fermion (HF) compounds and other materials. Recent theoretical attempts to classify QCPs in HF systems~\cite{Gegenwart2008,Si2010} rely on the knowledge of whether the $f$ electrons are itinerant or localized on both sides of a QCP~\cite{Si2006,Coleman2010}. The Fermi surfaces (FSs) with localized and itinerant $f$ electrons are commonly referred to as ``small'' and ``large'', respectively, as in the latter case the $f$ electrons effectively contribute to the FS. The FS topology is often determined from de Haas-van Alphen (dHvA) effect measurements, in which each oscillatory frequency is proportional to an extreme cross section of the FS in the direction perpendicular to the applied magnetic field. Therefore, ``small'' FSs are expected to give rise to low oscillatory frequencies, whereas ``large'' FSs should result in higher frequencies observed in such measurements. This was indeed observed in the family of Ce$M$In$_5$ ($M =$ Co, Ir, Rh) compounds. The topology of the main FSs of nonmagnetic CeCoIn$_5$~\cite{Settai2001,Hall2001,Shishido2002} and CeIrIn$_5$~\cite{Haga2001,Shishido2002} with itinerant $f$ electrons is essentially the same as that of the 4$f$-localized FSs of antiferromagnetic (AF) CeRhIn$_5$~\cite{Hall2001a,Shishido2002}. As expected, the size of the FSs of CeCoIn$_5$ and CeIrIn$_5$ is larger than that of the FSs of CeRhIn$_5$, resulting in higher dHvA frequencies observed in the former compounds~\cite{Shishido2002}. The dHvA effect measurements performed in CeRhIn$_5$ under pressure revealed an increase of all the dHvA frequencies above the critical pressure to suppress antiferromagnetism, which was naturally interpreted as the delocalization of the $f$ electrons~\cite{Shishido2005}.

In general, however, the FSs of HF materials with either itinerant or localized $f$ electrons are very complex, and the topology of the former is often entirely different from the latter. Moreover, even the size of the FSs in some HF compounds with itinerant $f$ electrons is smaller than that in their $f$-localized analogs. This is, for example, the case in paramagnetic (PM) CeRu$_2$Si$_2$ with itinerant $f$ electrons, in which the band-structure calculations predict a large hole sheet of the FS in the 14th band~\cite{Yamagami1993}. This FS is smaller than its counterpart calculated for LaRu$_2$Si$_2$~\cite{Yamagami1992}, which corresponds to CeRu$_2$Si$_2$ with localized $f$ electrons. The calculated FSs of LaRu$_2$Si$_2$~\cite{Yamagami1993} are in good agreement with those experimentally observed in ferromagnetic CeRu$_2$Ge$_2$ with localized $f$ electrons~\cite{King1991,Ikezawa1997,Yamagami1994}, in which some of the experimentally observed dHvA frequencies are indeed higher than the corresponding frequencies in CeRu$_2$Si$_2$ below the metamagnetic transition that occurs at about 8 T when the field is applied along the $c$ axis~\cite{Takashita1996}.

It can then be difficult to distinguish ``small'' and ``large'' FSs from the experimental results alone. Nonetheless, this can be achieved by comparing the experimentally obtained angular dependence of the dHvA frequencies with the results of band-structure calculations performed for both itinerant and localized $f$ electrons. In the simple case of Ce-based HF compounds with only one $f$ electron, this corresponds to the calculations for a particular Ce-based compound itself and its La-based analog without $f$ electrons. Such calculations in HF compounds are usually performed for a PM ground state. Then the comparison of the dHvA frequencies experimentally observed in magnetically ordered HF materials with band-structure calculations is not straightforward. Indeed, the AF order changes the topology of the FS. This might be approximated by a band-folding procedure where the PM FS is folded into a small Brillouin zone (BZ) based on a large magnetic unit cell. As a result, only small PM FSs located close to the center of the BZ survive in the AF state. On the other hand, new small FSs can be formed by magnetic order. Nonetheless, large PM FSs exceeding in size the magnetic BZ were observed in AF CeRhIn$_5$~\cite{Hall2001a,Shishido2002}, CeIn$_3$~\cite{Harrison2007}, Ce$_2$RhIn$_8$~\cite{Ueda2004}, and CePt$_2$In$_7$~\cite{Altarawneh2011}. Such an observation is possible due to a magnetic breakdown, which often takes place at a relatively low field in HF materials because of the small ordered moments typical for such systems. Then the BZ reconstruction by the magnetic order can be neglected.

The tetragonal HF compound CeRh\(_2\)Si\(_2\) with a moderate specific heat Sommerfeld coefficient \(\gamma = 22.8\)~mJ/K\(^2\)mol~\cite{Graf1997} orders antiferromagnetically  at \(T_{N1} = 36\)~K. Its magnetic structure changes again at \(T_{N2} = 24\)~K~\cite{Kawarazaki2000}. Both \(T_{N1}\) and \(T_{N2}\) are suppressed by pressure, $P$, and disappear at \(P_{c1} \sim 1.0\)~GPa and \(P_{c2} \sim 0.6\)~GPa, respectively~\cite{Kawarazaki2000,Araki2002}. Superconductivity was observed near the critical pressure \(P_{c1}\) required to completely suppress AF ordering~\cite{Movshovich1996,Araki2002}. When the magnetic field is applied along the crystallographic \(c\) axis, a complex two-step first-order metamagnetic (MM) transition to a polarized PM state occurs at about 26~T at low temperatures~\cite{Settai1997,Abe1998,Knafo2010,PalacioMorales2015}.

The dHvA oscillations in CeRh\(_2\)Si\(_2\) have been observed both at ambient pressure~\cite{Abe1998a, Araki1998} and under pressure, in which case the magnetic field was applied along the $a$ axis~\cite{Araki2001,Onuki2004}. At $P = 0$, numerous dHvA frequencies were observed. Most of these frequencies, not predicted by the band-structure calculations performed for a PM ground state~\cite{Araki2001}, originate from the band folding of the FSs in the AF state. Indeed, the low-temperature magnetic structure of CeRh\(_2\)Si\(_2\) suggests that its magnetic unit cell is eight times larger than the chemical one~\cite{Kawarazaki2000}. Correspondingly, the BZ in the AF state is reduced to one-eighth compared to that in the PM state. However, two very low frequencies, 44 and 137 T, detected around the $c$ axis correspond to small FS pockets predicted by the 4\(f\)-localized band-structure calculations. Furthermore, several much higher frequencies observed farther away from the $c$ axis are well accounted for by large FSs with localized 4\(f\) electrons. These frequencies are likely to result from a magnetic breakdown. The effective masses are rather small, of the order of a few bare electron masses. At \(P_{c2}\), some of the initially observed dHvA branches disappear, and new frequencies emerge. The change of the FS is even more drastic at \(P_{c1}\). All of the previously observed dHvA frequencies disappear and are replaced by three completely new branches. The new branches are well described by the 4\(f\)-itinerant band model~\cite{Araki2001,Onuki2004}. One of the effective masses increases dramatically upon approaching \(P_{c1}\)~\cite{Araki2001,Onuki2004}. Recently, quantum oscillations were observed in high field thermoelectric power measurements at $P = 0$~\cite{PalacioMorales2015}. The oscillations suddenly disappeared above the MM transition, suggesting a FS reconstruction.

In this Rapid Communication, we report dHvA effect measurements in CeRh$_2$Si$_2$ across its field-induced MM transition. Relatively low dHvA frequencies observed below the transition are replaced by much higher ones above it, implying a drastic change of the FS size. However, a comparison of the dHvA frequencies and their angular dependence above the transition with band-structure calculations suggests that the $f$ electrons remain localized in the field-induced polarized PM state.

\begin{figure}[htb]
\includegraphics[width=7.5cm]{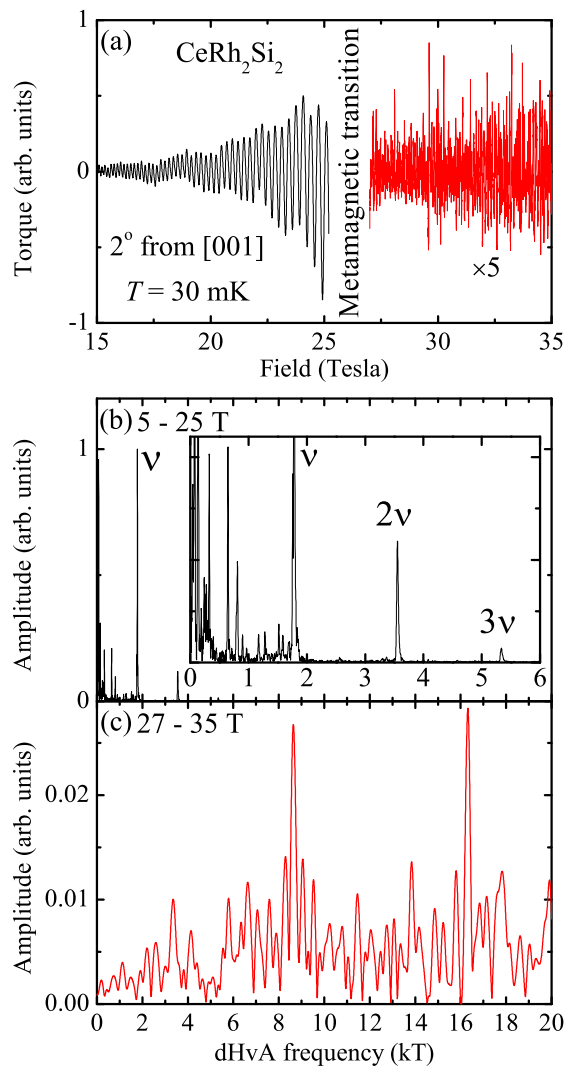}
\caption{\label{fig:dHvA} (a) dHvA oscillations and the corresponding FFT spectra (b) below and (c) above the MM transition. The inset shows a zoom of the low frequency part of the FFT spectrum below the transition.}
\end{figure}

A high quality single crystal of CeRh\(_2\)Si\(_2\) with a residual resistivity ratio of about 100 used in our study was grown by the Czochralski method in a tetra-arc furnace. The dHvA measurements were performed using a torque cantilever magnetometer mounted in a top-loading dilution refrigerator equipped with a low-temperature rotator. Magnetic fields up to 35 T were applied at different angles between the [001] and [100] directions. Due to a strong jump in magnetic torque at the MM transition, the angular range was limited to 8$^\circ$ off the $c$ axis.

We observed dHvA oscillations both below and above the MM transition, as shown in Fig.~\ref{fig:dHvA} for a magnetic field applied at 2$^\circ$ off the $c$ axis. The fast Fourier transform (FFT) spectrum of the dHvA oscillations below the MM transition [Fig.~\ref{fig:dHvA} (b)] is dominated by a peak $\nu$ at 1.8 kT and its harmonics in good agreement with previous lower field results~\cite{Abe1998a,Araki1998,Araki2001}. We did not observe any dHvA frequencies above 6 kT in the AF state.

Above the MM transition, the amplitude of the dHvA oscillations is much smaller. Nonetheless, two high dHvA frequencies of 8.6 and 16.3 kT are clearly seen in the FFT spectrum [Fig.~\ref{fig:dHvA} (c)]. These frequencies are much higher than the frequencies observed below the transition both in our own and previously reported measurements for the same orientation of the magnetic field~\cite{Abe1998a,Araki1998,Araki2001}. This implies a drastic modification of the FS across the MM transition in CeRh$_2$Si$_2$.

\begin{figure}[htb]
\includegraphics[width=7.5cm]{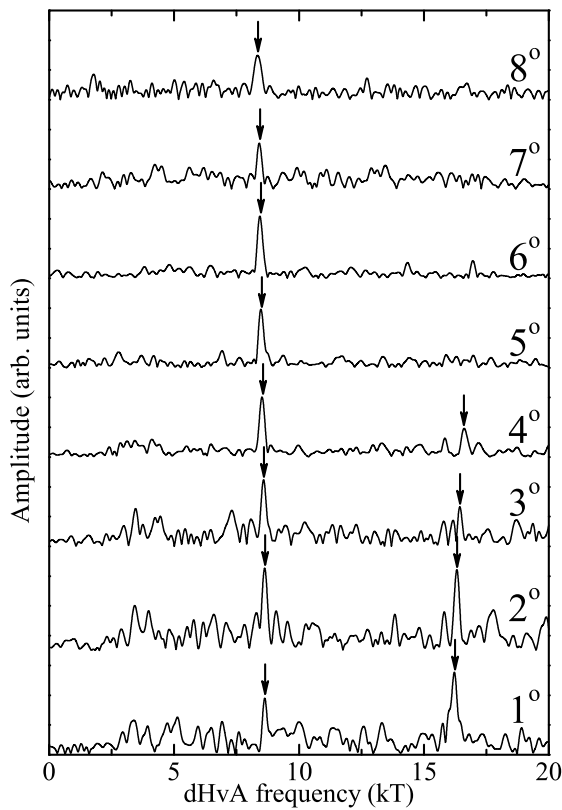}
\caption{\label{fig:FFT}FFT spectra of the dHvA oscillations above the MM transition at different angles between the magnetic field and the $c$ axis.}
\end{figure}

Figure~\ref{fig:FFT} shows the FFT spectra of the dHvA oscillations observed above the MM transition at different angles from the $c$ axis. The higher frequency increases with angle and is observed over a very limited angular range, up to 4$^\circ$. The lower frequency decreases with angle and is observed over the whole angular range of our measurements.

\begin{figure}[htb]
\includegraphics[width=7.5cm]{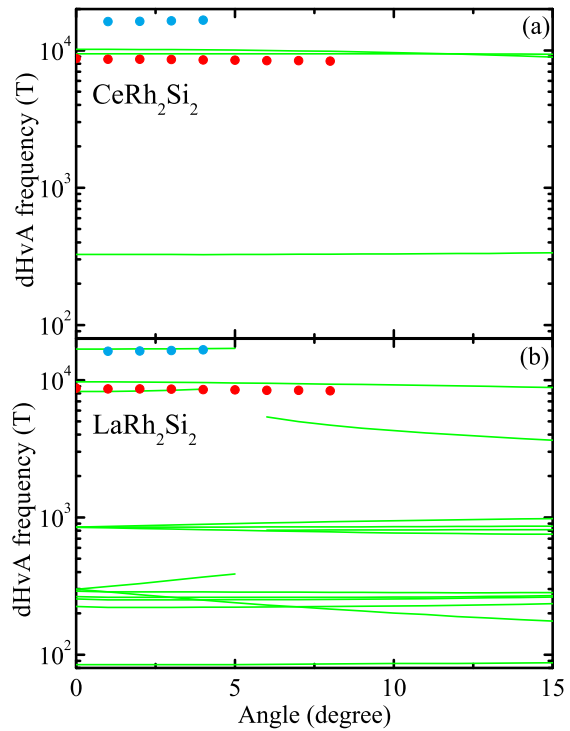}
\caption{\label{fig:AngDep}Angular dependence of the calculated dHvA frequencies (lines) for (a) CeRh$_2$Si$_2$ and (b) LaRh$_2$Si$_2$, together with the experimental results obtained in CeRh$_2$Si$_2$ above the MM transition (solid symbols).}
\end{figure}

\begin{figure*}[htb]
\includegraphics[width=15cm]{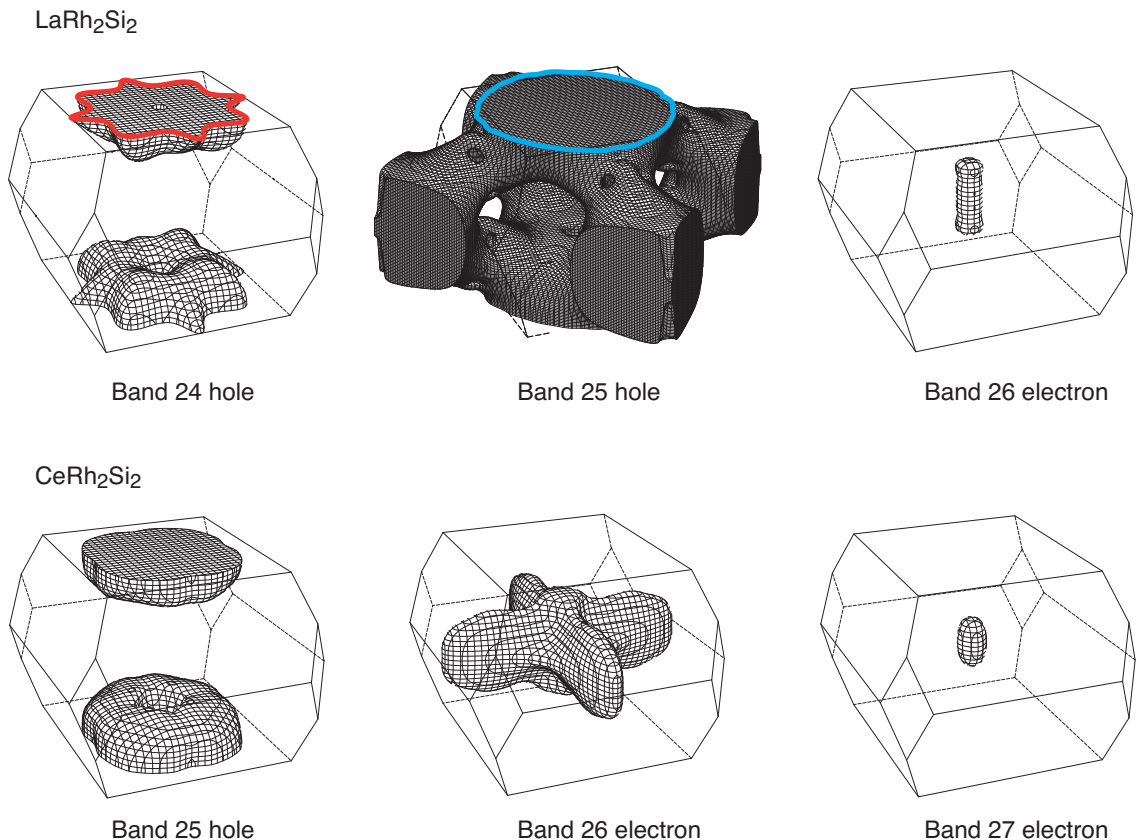}
\caption{\label{FS}Calculated FSs of LaRh$_2$Si$_2$ and CeRh$_2$Si$_2$. For LaRh$_2$Si$_2$, the orbits giving rise to the experimentally observed high dHvA frequencies are shown.}
\end{figure*}

In order to figure out whether the $f$ electrons are itinerant or localized above the MM transition, we compared the experimentally observed angular dependence of the dHvA frequencies with the results of band-structure calculations performed for both CeRh$_2$Si$_2$ (itinerant $f$ electron model) and LaRh$_2$Si$_2$ (localized $f$ electron model). In the latter case, the lattice parameters of CeRh$_2$Si$_2$ were used for the calculations. Figure~\ref{fig:AngDep} shows both measured and calculated dHvA frequencies, while the calculated FSs are shown in Fig.~\ref{FS}.

As can be seen in Fig.~\ref{fig:AngDep}, the angular dependence of the lower dHvA frequency observed above the transition can be fairly well accounted for by the results of the band-structure calculations performed for both itinerant and localized $f$ electrons. Therefore, this frequency alone is not sufficient to distinguish unambiguously which model better describes the experimental results obtained above the transition. On the other hand, there is no branch corresponding to the higher frequency in the calculations performed for CeRh$_2$Si$_2$. The calculations performed within the localized $f$ electron model, by contrast, predict a branch that agrees perfectly well with the observed higher frequency and its angular dependence. Even the angular range where this frequency exists is reproduced well in the calculations. A small difference between the absolute values of the experimentally observed and calculated dHvA frequencies is probably due to a strong volume expansion, $\Delta V/V \sim 1\times10^{-3}$, at the MM transition revealed by magnetostriction measurements~\cite{Naito2003}.

The comparison of the experimentally observed dHvA frequencies above the MM transition in CeRh$_2$Si$_2$ with the results of the band-structure calculations thus provides clear evidence that the $f$ electrons remain localized in the polarized PM state above the MM transition. According to the calculations performed for LaRh$_2$Si$_2$, the lower dHvA frequency observed at high field corresponds to an orbit originating from a flat starlike FS in band 24 (Fig.~\ref{FS}). The branch corresponding to the higher frequency originates from a complicated multiconnected FS in band 25 (Fig.~\ref{FS}). In both cases, the FSs have an unfavorable curvature factor, which accounts for a small amplitude of the dHvA oscillations observed above the MM transition.

In Ce-based HF compounds, a modification of the FS related to the delocalization of the $f$ electrons is theoretically expected to occur either exactly at a QCP or inside the magnetically ordered phase~\cite{Gegenwart2008,Si2010}.  For a pressure-induced QCP, such a modification was indeed observed in CeRhIn$_5$~\cite{Shishido2005}, CeIn$_3$~\cite{Settai2005}, and the compound in question, CeRh$_2$Si$_2$~\cite{Araki2001,Onuki2004} exactly at $P_c$. On the other hand, such a change is unlikely to occur at a field-induced QCP, as high magnetic fields polarize the electronic bands and therefore lead to the localization of the $f$ electrons. For example, the $f$ electrons in CeIn$_3$ are still localized above the N\'{e}el critical field of about 61~T~\cite{Harrison2007}. It is therefore not surprising that the $f$ electrons remain localized above the MM transition in CeRh$_2$Si$_2$, where a field-induced QCP ia avoided due to the first-order nature of the MM transition. Magnetization measurements in CeRh$_2$Si$_2$ reveal the magnetic moment of 1.5 $\mu_{\rm B}$/Ce above the MM transition~\cite{Settai1997} implying a strong magnetic polarization.

In CeRh$_2$Si$_2$, the change of the dHvA frequencies from low below the MM transition to high above it suggests a corresponding change of the FS size from small to large. However, this does not correspond to a change from the FS with localized $f$ electrons to the FS with itinerant ones. The drastic change of the experimentally observed FS size is rather due to the restoration of the crystallographic BZ in the polarized PM state.

In summary, we observed quantum oscillations both below and above the first-order MM transition from the AF to the polarized PM phase in CeRh$_2$Si$_2$. Low dHvA frequencies observed below the transition are found to give way to much higher ones above it. This suggests a drastic change of the FS size from small below the transition to large above it. However, the high frequencies observed above the transitions are best accounted for by the band-structure calculations performed for LaRh$_2$Si$_2$, implying that the $f$ electrons remain localized in the field-induced polarized PM state. Rather, a drastic change of the FS is caused by the modification of the BZ at the MM transition. This example of the FS reconstruction from small to large without delocalization of the $f$ electrons emphasizes the ambiguity of the commonly used simplistic conception of ``small'' and ``large'' FSs for localized and itinerant $f$ electrons, respectively. The FS modification we observe here is different from that reported to occur at the pressure-induced QCP, where the $f$ electrons do delocalize and become itinerant. From this point of view, it would be interesting to compare our results above the MM transition with the dHvA frequencies in the pressure-induced PM state. To this end, dHvA measurements under pressure are to be carried out with the magnetic field applied along the $c$ axis.

\begin{acknowledgments}
We thank Y. \={O}nuki, S. Araki, and M.T. Suzuki for useful discussions. KG acknowledges support from the DFG within GRK 1621. We acknowledge the support of the LNCMI-CNRS and the HLD-HZDR, members of the European Magnetic Field Laboratory (EMFL), the ANR-DFG grant ``Fermi-NESt'', and JSPS KAKENHI Grants number 15K05882, 15H05884, 15H05886, 15K21732 (J-Physics).
\end{acknowledgments}

\bibliography{CRSdHvA}

\end{document}